\documentclass[11pt,a4paper]{article}
\usepackage{amsmath}
\usepackage{amssymb}
\usepackage{amscd}
\usepackage[dvips]{graphicx}

\numberwithin{equation}{section}

\begin{document}

\newtheorem{thm}{Theorem}[section]
\newtheorem{cor}{Corollary}[section]
\newtheorem{lem}{Lemma}[section]
\newtheorem{prop}{Proposition}[section]
\newtheorem{defn}{Definition}[section]
\newtheorem{exas}{Examples}[section]
\newtheorem{exam}{Example}[section]
\newtheorem{counterexam}{Counterexample}[section]
\newtheorem{rem}{Remark}[section]
\newtheorem{ques}{Question}[section]
\newtheorem{conj}{Conjecture}[section]
\numberwithin{equation}{section}


\title{Results on the solutions of maximum weighted Renyi entropy problems }

\author{Salimeh Yasaei Sekeh}
\date{}
\maketitle

\begin{abstract}

In this paper, following standard arguments, the maximum Renyi entropy problem for the weighted case is analyzed. We verify that under some constrains on weight function, the Student-$r$ and Student-$t$ distributions maximize the weighted Renyi entropy. Furthermore, an extended version of the Hadamard inequality is derived.

\end{abstract}


\def\wt{\widetilde}
\def\fB{\mathfrak B}\def\fM{\mathfrak M}\def\fX{\mathfrak X}
 \def\cB{\mathcal B}\def\cM{\mathcal M}\def\cX{\mathcal X}
\def\be{\mathbf e}
\def\bu{\mathbf u}\def\bv{\mathbf v}\def\bx{\mathbf x} \def\by{\mathbf y} \def\bz{\mathbf z}
\def\om{\omega} \def\Om{\Omega}
\def\bbP{\mathbb P} \def\hw{h^{\rm w}} \def\hwi{{h^{\rm w}}}
\def\beq{\begin{eqnarray}} \def\eeq{\end{eqnarray}}
\def\beqq{\begin{eqnarray*}} \def\eeqq{\end{eqnarray*}}
\def\rd{{\rm d}} \def\Dwphi{{D^{\rm w}_\phi}}
\def\BX{\mathbf{X}}\def\Lam{\Lambda}\def\BY{\mathbf{Y}}
\def\BZ{\mathbf{Z}} \def\BN{\mathbf{N}}

\def\tM{\tilde M}
\def\mwe{{D^{\rm w}_\phi}}
\def\DwPhi{{D^{\rm w}_\Phi}} \def\iw{i^{\rm w}_{\phi}}
\def\bE{\mathbb{E}}
\def\1{{\mathbf 1}} \def\fB{{\mathfrak B}}  \def\fM{{\mathfrak M}}
\def\diy{\displaystyle} \def\bbE{{\mathbb E}} \def\bu{\mathbf u}
\def\BC{{\mathbf C}} \def\lam{\lambda} \def\bbB{{\mathbb B}}
\def\bbR{{\mathbb R}}\def\bbS{{\mathbb S}}
 \def\bmu{{\mbox{\boldmath${\mu}$}}}
 \def\bPhi{{\mbox{\boldmath${\Phi}$}}}  \def\bPi{{\mbox{\boldmath{$\Pi$}}}}
  \def\btheta{{\mbox{\boldmath${\theta}$}}}
 \def\bbZ{{\mathbb Z}} \def\fF{\mathfrak F}\def\bt{\mathbf t}\def\B1{\mathbf 1}
\def\hwphi{h^{\rm w}_{\phi}}
\def\BW{\mathbf{W}} \def\bw{\mathbf{w}}
\def\beal{\begin{array}{l}}
\def\beac{\begin{array}{c}}
\def\beacl{\begin{array}{cl}}
\def\ena{\end{array}}
\def\WBJ{\mathbf{J}^{\rm w}_{\phi}}
\def\BS{\mathbf{S}}
\def\BK{\mathbf{K}}
\def\BB{\mathbf{B}}
\def\wtD{{\widetilde D}}

\def\mwe{{D^{\rm w}_\phi}}
\def\DwPhi{{D^{\rm w}_\Phi}} \def\iw{i^{\rm w}_{\phi}}
\def\bE{\mathbb{E}}
\def\1{{\mathbf 1}} \def\fB{{\mathfrak B}}  \def\fM{{\mathfrak M}}
\def\diy{\displaystyle} \def\bbE{{\mathbb E}} \def\bu{\mathbf u}
\def\BC{{\mathbf C}} \def\lam{\lambda}
\def\bbB{{\mathbb B}} \def\bbM{{\mathbb M}}
\def\bbR{{\mathbb R}}\def\bbS{{\mathbb S}}
\def\blam{{\mbox{\boldmath${\lambda}$}}}
\def\bmu{{\mbox{\boldmath${\mu}$}}} \def\bta{{\mbox{\boldmath${\eta}$}}}
\def\bzeta{{\mbox{\boldmath${\zeta}$}}}
 \def\bPhi{{\mbox{\boldmath${\Phi}$}}}  \def\bPi{{\mbox{\boldmath{$\Pi$}}}}
 \def\bbZ{{\mathbb Z}} \def\fF{\mathfrak F}\def\bt{\mathbf t}\def\B1{\mathbf 1}
\def\hwphi{h^{\rm w}_{\phi}}
\def\BT{{\mathbf T}} \def\BW{\mathbf{W}} \def\bw{\mathbf{w}}
\def\beal{\begin{array}{l}}
\def\beac{\begin{array}{c}}
\def\beacl{\begin{array}{cl}}
\def\ena{\end{array}}
\def\WBJ{\mathbf{J}^{\rm w}_{\phi}}
\def\BS{\mathbf{S}}
\def\BK{\mathbf{K}}
\def\tL{\mathbf{L}}
\def\BB{\mathbf{B}}
\def\vphi{{\varphi}}
\def\rw{{\rm w}}
\def\bZ{\mathbf Z} \def\b0{\mathbf{0}}
\def\wtf{{\widetilde f}} \def\wtg{{\widetilde g}} \def\wtG{{\widetilde G}}
\def\vphi{\varphi}
\def\rT{{\rm T}}
\def\tA{{\tt A}} \def\tB{{\tt B}} \def\tC{{\tt C}} \def\tI{{\tt I}} \def\tJ{{\tt J}} \def\tK{{\tt K}}
\def\tL{{\tt L}} \def\tP{{\tt P}} \def\tQ{{\tt Q}} \def\tS{{\tt S}}
\def\fB{\mathfrak B}\def\fM{\mathfrak M}\def\fX{\mathfrak X}
 \def\cB{\mathcal B}\def\cM{\mathcal M}\def\cX{\mathcal X}
\def\bu{\mathbf u}\def\bv{\mathbf v}\def\bx{\mathbf x}\def\by{\mathbf y}
\def\om{\omega} \def\Om{\Omega}
\def\bbP{\mathbb P} \def\hw{{h^{\rm w}}} \def\hwphi{{h^{\rm w}_\phi}}
\def\beq{\begin{eqnarray}} \def\eeq{\end{eqnarray}}
\def\beqq{\begin{eqnarray*}} \def\eeqq{\end{eqnarray*}}
\def\rd{{\rm d}} \def\Dwphi{{D^{\rm w}_\phi}}
\def\BX{\mathbf{X}}
\def\hwphiii{{h^{\rm w}_{\phi_1\otimes\phi_2\otimes\dots\otimes \phi_n}}} \def\hwphii{{h^{\rm w}_{\phi_1\otimes\phi_2}}}
\def\mwe{{D^{\rm w}_\phi}}
\def\DwPhi{{D^{\rm w}_\Phi}} \def\iw{i^{\rm w}_{\phi}}
\def\bE{\mathbb{E}}
\def\1{{\mathbf 1}} \def\fB{{\mathfrak B}}  \def\fM{{\mathfrak M}}
\def\diy{\displaystyle} \def\bbE{{\mathbb E}}
\def\fB{\mathfrak B}\def\fM{\mathfrak M}\def\fX{\mathfrak X}
 \def\cB{\mathcal B}\def\cM{\mathcal M}\def\cX{\mathcal X}
\def\bu{\mathbf u}\def\bv{\mathbf v}\def\bx{\mathbf x}\def\by{\mathbf y}
\def\om{\omega} \def\Om{\Omega}
\def\bbP{\mathbb P} \def\hw{h^{\rm w}} \def\hwi{{h^{\rm w}}}
\def\beq{\begin{eqnarray}} \def\eeq{\end{eqnarray}}
\def\beqq{\begin{eqnarray*}} \def\eeqq{\end{eqnarray*}}
\def\rd{{\rm d}} \def\Dwphi{{D^{\rm w}_\phi}}
\def\BX{\mathbf{X}}\def\Lam{\Lambda}\def\BY{\mathbf{Y}}

\def\mwe{{D^{\rm w}_\phi}}
\def\DwPhi{{D^{\rm w}_\Phi}} \def\iw{i^{\rm w}_{\phi}}
\def\bE{\mathbb{E}}
\def\1{{\mathbf 1}} \def\fB{{\mathfrak B}}  \def\fM{{\mathfrak M}}
\def\diy{\displaystyle} \def\bbE{{\mathbb E}} \def\bu{\mathbf u}
\def\BC{{\mathbf C}} \def\lam{\lambda} \def\bbB{{\mathbb B}}
\def\bbR{{\mathbb R}}\def\bbS{{\mathbb S}} \def\bmu{{\mbox{\boldmath${\mu}$}}}
 \def\bPhi{{\mbox{\boldmath${\Phi}$}}}
 \def\bbZ{{\mathbb Z}} \def\fF{\mathfrak F}\def\bt{\mathbf t}\def\B1{\mathbf 1}
\def\UX{\underline{X}}
\def\ux{\underline{x}}
\def\Hw{H^{\rm w}}
\def\OY{\overline{Y}} \def\oy{\overline{y}}
\def\OY{\overline{Y}} \def\oy{\overline{y}} \def\omu{\overline{\mu}} \def\OSigma{\overline{\Sigma}}
\def\bbS{\mathbb{S}}
\def\bF{\overline{F}}
\def\ew{\mathcal{E}^{\rm w}_{\vp}}
\def\vp{\varphi} \def\sw{\mathcal{S}^{\rm w}}
\def\inr{\int\limits_{\bbR_+^n}} \def\Dwvp{D^{\rm w}_{\varphi}} \def\hwvp{h^{\rm w}_{\varphi}}
\let\Dwphi\Dwvp
\let\hwphi\hwvp

\section{Introduction}

It is well-known that entropy has been widely played an important role in optimization problems which preserve varies applications in areas of computer vision, communication transmission, medical and so on presented in the literature. Thus, studying the entropy maximizing distributions became a principal object in information theory for understanding the Shannon entropy optimization, and later extended versions of problems such as Renyi and Tsallis entropies maximization. See \cite{MRS, Ka, Z, CHV, VCH, JV}.

In 1968-1971, subsequently followed by Shannon entropy concept, the initial definition of {\it weighted entropy} was illustrated in \cite{BG, G}. Following the weighted progress, recently further results in \cite{C, SY, SuYS, SuYSSt, SYK, SuStKel} with a number of theoretical suggestions have been established. As a kind of fundamental reference, in \cite{SY} the maximization of weighted entropy and its consequences were discussed as well.

For given function $\bx\in\bbR^n\mapsto\vp (\bx)\geq 0$, and an random vector (RV)
$\BX\in\bbR^n$, with a joint probability density function (PDF) $f$,
the weighted entropy (WE) of $\BX$ (or $f$) with  weight function (WF) $\vp$ is defined by
\beq\label{eq:1.1}\hwphi (\BX)=\hwphi (f) =-\int_{\bbR^n}\vp (\bx )f(\bx )\log\,f(\bx)\;\rd \bx=-\bbE_{\BX}(\vp\log\,f)\eeq
whenever the integral $\diy\int_{\bbR^n}\vp (\bx )f(\bx )\Big(1\vee|\log\,f(\bx)|\Big)\rd \bx<\infty$. (A
standard agreement $0=0\cdot\log\,0=0\cdot\log\,\infty$ is adopted throughout the paper). Furthermore, for two functions, $\bx\in\bbR^n\mapsto f(\bx)\geq 0$ and $\bx\in\bbR\mapsto g(\bx)\geq 0$,
the relative WE of $g$ relative to $f$ with WF $\vp$ is defined by
\beq\label{eq:1.2}
\Dwphi (f\|g)=\int_{\bbR^n}\vp (\bx )f(\bx )\log\frac{f(\bx )}{g(\bx )}\rd \bx.
\eeq
When $\vp\equiv 1$ the relative WE yields the Kullback-leibler divergence. Searching the weighted determinant inequalities as direct extension for the standard forms to non-constant weight function we refer the reader once more to \cite{BG, C, SY}. However, as consequences of those generalizations a number of new bounds in terms of determinants of positive definite matrices, by applying Gaussian WEs, has been given. As one step further, the author proposed another general form of the WE by introducing the {\it weighted Renyi entopy}, where in spite of standard case \cite{R}, in particular $p\rightarrow 1$ literally does not intend to the WE but a proportion of it, see \cite{S}.

\begin{defn}\label{Def:1.1}
The $p$-th weighted Renyi entropy (WRE) of a RV $\BX$ with PDF $f$ in $\bbR^n$, given WF $\vp$ and for $p>0,p\neq 1$, is defined by
\beq h^{\rm w}_{\vp,p}(\BX):=h^{\rm w}_{\vp,p}(f)=\diy\frac{1}{1-p}\log \int_{\bbR^n} \vp(\bx) f^p(\bx)\rd \bx.\eeq
Observe that if $\varphi\equiv 1$, the WRE, $h^{\rm w}_{\vp,p}(f)$, becomes the known Renyi entropy, denoted by $h_p(f)$, cf. \cite{R}. Since $\vp\geq 0$, it can be checked that like Renyi entropy, for $0<p<1$ the WRE is a concave function whereas when $p>1$ we can not make a similar statement.
\end{defn}
Observe that
\beq\label{Eq:1.5} \lim\limits_{p\rightarrow 1}h^{\rm w}_{\vp,p}(f)=h^{\rm w}_{\vp,1}(f)=\diy \frac{h^{\rm w}_{\vp}(f)}{\bbE_f[\vp]}.\eeq
On the other words as ${p\rightarrow 1}$ the WRE does not intend to the weighted entropy (WE) precisely, see \cite{CT, KS2, SYK}.
Note that both $h^{\rm w}_{\vp,p}(f)$ is a continuous function in $p$.\\

Next, extending the standard notions, the {\bf relative} $p$-th weighted Renyi power (WRP) of $f$ and $g$ was proposed: for $p>0,p\neq 1$ and given WF $\bx\in\bbR^n\mapsto\vp(\bx)\geq 0$
\beq\label{RPWRP} N^{\rm w}_{\vp,p}(f,g)=\diy \frac{\bigg(\diy\int_{\bbR^n} \vp\; g^{p-1}f \rd \bx\bigg)^{1/(1-p)}\bigg(\diy\int_{\bbR^n}\vp\; g^p \rd \bx\bigg)^{1/p}}{\bigg(\diy\int_{\bbR^n}\vp\; f^p\rd \bx\bigg)^{1\big/p(1-p)}}.\eeq
And more generally, for given two functions $f$ and $g$ we employed the relative $p$-th WRP in order to define the relative $p$-th weighted Renyi entropy (WRE) of $f$ and $g$ by
\beq D^{\rm w}_{\vp,p}(f\|g)=\log N^{\rm w}_{\vp,p}(f,g).\eeq
Hence one can write
\beq \label{Eq:1.7}D^{\rm w}_{\vp,p}(f\|g)=\diy\frac{1}{1-p}\log\Big(\int_{\bbR^n}\varphi\;g^{p-1}\;f\;\rd \bx\Big)+\frac{1-p}{p} h^{\rm w}_{\varphi,p}(g)-\frac{1}{p} h^{\rm w}_{\varphi,p}(f). \eeq
Particularly Eq. (\ref{Eq:1.7}) yields
\beq\label{Eq:1.1} D^{\rm w}_{\vp,1}(f,g)=\lim\limits_{p\rightarrow 1}D^{\rm w}_{\vp,p}(f,g)=\diy \frac{\Dwphi (f\|g)}{\bbE_f[\vp]}.\eeq

The reflection of the importance of maximum entropy problems, following the terminology used in \cite{CHV, JV}, features the present paper. We intend to study the Renyi entropy maximizing distributions in weighted case for understanding the advantages and limitations of such extended version of entropy maximizing method. In addition the results presented in the current paper obtains parallels to some of the properties involving determinants of matrices by using the weighted Renyi entropy maximizer. Thus this work is organized first by reviewing the multivariate Student-t and Student-r distributions. Furthermore, we show that under some constrains in terms of the WF, these distributions maximize the WRE for cases $p<1$ and $p>1$ distinctly. As a primary consequence, we consider the so-called weighted Hadamard inequality, \cite{SY} and extend the bound by taking into account the Pearson II and VII PDFs based on the chain rule for the WRE.

Let us begin with a transparent result as a direct application of H\"{o}lder inequality below. In fact, this is n-dimensional version of Theorem 1.1 in \cite{YS}, hence the proof is omitted.
\begin{lem}\label{lem1}
For $p>0$  and PDFs $f$, $g$ and given WF $\vp$, assume if $p=1$, $\bbE_f[\vp]\geq \bbE_g[\vp]$ holds. The relative $p$-th WRE $D^{\rm w}_{\vp,p}(f\|g)\geq 0$. Equality occurs when $f\equiv g$ almost everywhere.
\end{lem}
Reviewing the Renyi maximizing densities, Student-t (Pearson type VII) and Student-r (Pearson type II) distributions, we use the same notation as in \cite{JV}, $g_{p,\BC}$, and establish the following definition. For some of their properties the reader may refer to \cite{CHV}. Note that we will set $x_+=\max\{x,0\}$.
\begin{defn}\label{def:gc}
Define the n-dimensional PDF $g_{p,\BC}$ as
\beq\label{def.gc} g_{p,\BC}(\bx)=\left\{\begin{array}{ll} A_p\Big(1+(1-p)\beta\;\bx^T\BC^{-1}\bx\Big)^{\frac{1}{p-1}}_+,\quad p>n\big/(n+2),& p\neq 1,\\
\big((2\pi)^n{\rm det}\;\BC\big)^{-1/2}\exp\Big\{-\frac{1}{2} \bx^T\BC^{-1}\bx\Big\},& p=1.\end{array}\right.\eeq
where
\beq\label{beta} \beta=\diy\frac{1}{2p-n(1-p)},\quad \hbox{and}\eeq
\beqq A_p=\left\{\begin{array}{cc} \bigg(\Gamma(\frac{1}{1-p})\big(\beta(1-p)\big)^{n/2}\bigg)\Big/\bigg(\Gamma(\frac{1}{1-p}-\frac{n}{2})\pi^{n/2}({\rm det}\;\BC)^{1/2}\bigg), & \frac{n}{n+2}<p<1, \\
\bigg(\Gamma(\frac{p}{p-1}+\frac{n}{2})\big(\beta(p-1)\big)^{n/2}\bigg)\Big/\bigg(\Gamma(\frac{p}{p-1})\pi^{n/2}({\rm det}\;\BC)^{1/2}\bigg), & p>1.                                   \end{array}\right.\eeqq
Here $\Gamma$ stands the Gamma function. For brevity we will use $\bbS_{p,\BC}$ for the support of PDF $g_{p,\BC}$, hence if $p<1$, $\bbS_{p,\BC}=\bbR^n$ and for $p>1$, $\bbS_{p,\BC}=\{\bx,\; \bx^T\BC^{-1}\bx\leq 2p\big/(p-1)+n\}$.
\end{defn}

We briefly study the Pearson's type II and VII multivariate distributions by referring to \cite{Zo}, which we will recall them throughout the paper. From now on, because of homogeneity, we shall use only Pearson's type II and VII names for these kind of PDFs.

Let $\bbS=\{\bx\in\bbR^n, \bx^T\bx\leq 1\}$, with $\bbR^n$ being the $n$-dimensional Euclidian space. Then the Pearson's type II and VII with parameter $\mu$, denoted by $f_{II}(\bx;\mu)$, $f_{VII}(\bx;\mu)$, are defined as follows respectively:
\beq\begin{array}{cl} f_{II}(\bx;\mu)=\diy \frac{\Gamma(n/2+\mu+1)}{\pi^{n/2}\Gamma(\mu+1)}\big(1-\bx^T\bx\big)^{\mu},\quad \bx\in\bbS,\;\;\mu>-1,\\
\\
f_{VII}(\bx;\mu)=\diy \frac{\Gamma(\mu)}{\pi^{n/2}\Gamma(\mu-n/2)}\big(1+\bx^T\bx\big)^{-\mu},\quad \bx\in\bbR^n,\;\;\mu>n/2.\end{array}\eeq
Furthermore, for $n/(n+2)<p<1$, $q>(1-p)n/2$ set
\beq\label{varpi*} {\varpi}^*_{n}(p,q)=\diy\frac{\Gamma^q\big(1/(1-p)\big)\big(\beta(1-p)\big)^{n(q-1)/2}\Gamma\big(q/(1-p)-n/2\big)}
{\Gamma^q\big(1/(1-p)-n/2\big)\pi^{n(q-1)/2}\;\Gamma\big(q/(1-p)\big)},\eeq
and for $p>1$, $q>0$ denote
\beq\label{varpi} \varpi_n(p,q)=\diy \frac{\Gamma^q\big(p/(1-p)+n/2\big)\big(\beta(p-1)\big)^{n(q-1)/2}\Gamma\big(q/(p-1)+1\big)}
{\Gamma^q\big(p/(p-1)\big)\pi^{n(q-1)/2}\Gamma\big(n/2+q/(p-1)+1\big)}.\eeq
Here $\beta$ is as before (\ref{beta}). Also note that we will use ${\varpi}^*_{n}(p),\;{\varpi}_{n}(p)$ when in (\ref{varpi*}) and (\ref{varpi}) $p=q$.
Accordingly, let $\BY=(Y_1,\dots,Y_n)$, $\BY^*=(Y^*_1,\dots,Y^*_n)$ be RVs with PDFs Pearson's type II and VII, i.e. $\BY\sim f_{II}(.\; ; q/(p-1))$, $\BY^*\sim f_{VII}(.\; ;q/(1-p))$. We then introduce for $ n/(n+1)<p<1$, $q>(1-p)n/2$
\beq\label{Def:1.12} \begin{array}{l}
\rho^*:=\rho^*_{\vp,p}(\BY^*)=\vp\bigg(\Big\{2\big(p/(1-p)-n/2\big)\Big\}^{1/2}\BC^{1/2}\BY^*\bigg),\;\;\alpha^*_{\vp,p}(\BC)=\diy \bbE_{f_{VII}}\left[\rho^*_{\vp,p}(\BY^*)\right],\\
\hbox{and for}\;p>1,\;q>0\\
\rho:=\rho_{\vp,p}(\BY)=\vp\bigg(\Big\{2\big(p/(p-1)+n/2\big)\Big\}^{1/2}\BC^{1/2}\BY\bigg),\;\;\alpha_{\vp,p}(\BC)=\diy \bbE_{f_{II}}\left[\rho_{\vp,p}(\BY)\right].\end{array} \eeq
In addition, it is worthwhile to mention that in particular choice $\vp(\bx)=\prod\limits_{i=1}^n \vp_i(x_i)$ where $x_i\in\bbR\mapsto\vp_i(x_i)\geq 0$, the expression (\ref{Def:1.12}) takes the forms
\beq\label{Def:1.13} \begin{array}{l}
\alpha^*_{\vp,p}(\BC)=\diy \bbE_{f_{VII}}\left[\prod\limits_{i=1}^n\vp_i\bigg(\Big\{2\big(p/(1-p)-n/2\big)\Big\}^{1/2}\sum\limits_{j=1}^n Y^*_j\;C_{ij}^{1/2}\bigg)\right],\quad n/(n+1)<p<1,\\
\\
\alpha_{\vp,p}(\BC)=\diy \bbE_{f_{II}}\left[\prod\limits_{i=1}^n\vp_i\bigg(\Big\{2\big(p/(p-1)+n/2\big)\Big\}^{1/2}\sum\limits_{j=1}^n Y_j\;C_{ij}^{1/2}\bigg)\right],\quad p>1.\end{array} \eeq

Going back to the Definition \ref{def:gc}, we continue here the section by establishing the Weighted Renyi entropy for Renyi entropy maximizer and given WF $\vp$. Regarding the Pearson distributions suppose that $\BX\sim g_{p,\BC}$, therefore for  $n/(n+2)<p<1$, $q>(1-p)n/2$ one gets
\beq\label{WREgc}\begin{array}{l}
h^{\rm w}_{\vp,q}(g_{p,\BC})=\diy\frac{1}{1-q}\log \varpi^*_n(p,q)+\frac{1}{2}\log {\rm det}\;\BC+\frac{1}{1-q}\log \alpha^*_{\vp,p}(\BC).\end{array}\eeq
Moreover, if $p>1$, $q>0$ one obtains
\beq \label{WREgc2}\begin{array}{l} h^{\rm w}_{\vp,q}(g_{p,\BC})=\diy\frac{1}{1-q}\log \varpi_n(p,q)+\frac{1}{2}\log {\rm det}\;\BC+\frac{1}{1-q}\log \alpha_{\vp,p}(\BC).\end{array}\eeq

\section{Maximum weighted Renyi entropy}
As we said in the introduction, one of our goal in this work is to analyze the maximum weighted Renyi entropy. Precisely, following standard arguments, we first extend the result of Proposition 1.3, \cite{JV}, to exponents of the maximum WRE for RV $\BX$.
\begin{thm}\label{thm1}
For given $p>n/(n+2)$, consider RV $X$ with PDF $f$, mean $\b0$ and positive definite symmetric covariance matrix $\BC$. Let $\bx\in\bbR^n\mapsto\vp(\bx)$ be a given positive WF.  Define matrices $n\times n$, $\Psi^f=\big(\psi^f_{ij}\big)$ and $\Psi^{g}=\big(\psi^{g}_{ij}\big)$ where
\beqq \psi^f_{ij}=\diy\int_{\bbR^n}x_ix_j\vp(\bx)f(\bx)\;\rd \bx, \quad \psi^g_{ij}=\diy\int_{\bbR^n}x_ix_j\vp(\bx)g_{p,\BC}(\bx)\;\rd \bx.\eeqq
Assume that for $0<p<1(p>1)$
\beq\label{cond1}\diy\int_{\bbS_{p,\BC}}\vp(\bx)\Big[f(\bx)-g_{p,\BC}(\bx)\Big]\rd \bx+(1-p)\beta\;{\rm tr}\Big[\BC^{-1}\big(\Psi^f-\Psi^g\big)\Big] \leq (\geq) 0,\eeq
is fulfilled or consider the WF $\vp$ obeys
\beq\label{cond2}\begin{array}{l}
\diy\int_{\bbR^n} \vp(\bx)\Big[f(x)-g_{1,\BC}(\bx)\Big]\rd \bx\geq 0,\quad \hbox{and}\\
\diy\int_{\bbR^n} \vp(\bx)\log g_{1,\BC}(\bx)\Big[f(\bx)-g_{1,\BC}(\bx)\Big]\rd \bx\geq 0.\end{array}\eeq
Then under constrain (\ref{cond1})
\beq\label{maxWRE} h^{\rm w}_{\vp,p}(f)\leq h^{\rm w}_{\vp,p}(g_{p,\BC}), \quad p\neq 1,\eeq
holds and under (\ref{cond2}) one has
\beq \diy h^{\rm w}_{\vp}(f)\leq\diy h^{\rm w}_{\vp}(g_{1,\BC}),\eeq
with equality iff $f\equiv g_{p,\BC}$ almost everywhere. In fact the case $p=1$ literally illustrates the corresponding result in Example 3.2 cf. \cite{SY}.
\end{thm}
\vspace{0.5 cm}

{\bf Proof:}\;Using the Definition \ref{def:gc}, for $0<p<1(p>1)$ one can write
\beq\label{Eq:1.13} \begin{array}{l}
\diy\int_{\bbS_{p,\BC}}\vp(\bx)g_{p,\BC}^{p-1}(\bx)f(\bx)\rd \bx=\diy A_p^{p-1} \diy\int_{\bbS_{p,\BC}}\vp(\bx)\Big(1+(1-p)\beta \bx^T\BC^{-1}\bx\Big) f(\bx)\rd \bx\\
\qquad\qquad\leq(\geq)  \diy A_p^{p-1} \diy\int_{\bbS_{p,\BC}}\vp(\bx)\Big(1+(1-p)\beta \bx^T\BC^{-1}\bx\Big) g_{p,\BC}(\bx)\rd \bx\\
\qquad\qquad= \diy\int_{\bbS_{p,\BC}}\vp(\bx) g_{p,\BC}^{p}(\bx)\rd \bx. \end{array}\eeq
In expression (\ref{Eq:1.13}), the inequality is emerged from (\ref{cond1}), and in case $p=1$ it becomes
\beq \diy\int_{\bbS_{p,\BC}}\vp(\bx)f(\bx)\log g_{1,\BC}(\bx)\rd \bx\geq \diy\int_{\bbS_{p,\BC}}\vp(\bx)g_{1,\BC}(\bx)\log g_{1,\BC}(\bx)\rd \bx.\eeq
Now recall Lemma \ref{lem1}. Therefore owing to (\ref{Eq:1.7}) one yields
\beq\label{eq:2.7}\begin{array}{ccl} 0\leq D^{\rm w}_{\vp,p}(f\|g_{p,\BC})&=&\diy\frac{1}{1-p}\log\Big(\int_{\bbR^n}\varphi\;g_{p,\BC}^{p-1}\;f\;\rd \bx\Big)+\frac{1-p}{p} h^{\rm w}_{\varphi,p}(g_{p,\BC})-\frac{1}{p} h^{\rm w}_{\varphi,p}(f)\\
&\leq& \frac{1}{p}\bigg(h^{\rm w}_{\varphi,p}(g_{p,\BC})-h^{\rm w}_{\varphi,p}(f)\bigg).\end{array}\eeq
Likewise, the proof in case $p=1$ follows by repeating verbatim in the n-dimensional setting in Example 3.2 from \cite{SY}. $\quad$ $\blacksquare$

\begin{rem}
Considering arguments in \cite{CHV}, for given WF $\vp$, we introduce the non-symmetric directed divergence measure (see \cite{Cs}, \cite{AS}) for weighted case by
\beq D^{\rm w}_{\vp,p}(f\|g)={\rm sign}(p-1)\diy\int_{\bbS_{p,\BC}}\vp(\bx)\bigg(\frac{f^p(x)}{p}+\frac{p-1}{p} g^p(\bx)-f(\bx)g^{p-1}(\bx)\bigg)\rd \bx. \eeq
Going back to (\ref{Eq:1.13}), with the same strategy as Theorem \ref{thm1}, one deduces that under condition (\ref{cond1}), $D^{\rm w}_{\vp,p}(f\|g)\geq0$. This assertion implies (\ref{maxWRE}), which can be considered as alternative proof for Theorem \ref{thm1}.
\end{rem}

In sequel, in terms of a multivariate RV $\BX$, let us introduce
\beq\label{def:eta} \begin{array}{l} \eta^*_{\vp,p}(\mu)=\bbE_{f_{VII}}\Big[\vp\;\log\big(1+\BX^T\BX\big)\Big],\quad n/(n+2)<p<1,\\
\eta_{\vp,p}(\mu)=\bbE_{f_{II}}\Big[\vp\;\log\big(1-\BX^T\BX\big)\Big],\quad\quad p>1.\end{array}\eeq
where $\mu=p/(p-1)$. Using results obtained in \cite{Z}, next we shall follow the analogue methodology and apply the generalized spherical coordinate transformation. Therefore we able to establish the explicit quantities for various forms of the WF $\vp$. Let us begin with $\vp(\bx)=\bx^T\bx$, so we compute
\beq\label{eq2:2.9}\begin{array}{ccl} \eta^*_{\vp,p}(\mu)&=&\diy \frac{\Gamma(\mu)}{\pi^{n/2}\Gamma(\mu-n/2)}\int_{\bbR^n}\bx^T\bx\;\big(1+\bx^T\bx\big)^{-\mu}\log\big(1+\bx^T\bx\big)\rd\bx\\
\\
&=&\diy\frac{2\Gamma(\mu)}{\Gamma(n/2)\Gamma(\mu-n/2)}\int_0^\infty r^{n+1}(1+r^2)^{-\mu}\log(1+r^2)\rd r\\
\\
&=&\diy\frac{n}{2\mu-n-2}\Big\{\Psi\big(\mu\big)-\Psi\big(\mu-n/2-1\big)\Big\},\quad \mu>n/2+1.\end{array}\eeq
Here $\Psi(t)=\diy\frac{\rd}{\rd t}\log \Gamma(t)$. The last line in (\ref{eq2:2.9}) is derived by differentiating the Beta function defined by the following integral with respect to $\beta$:
\beqq B(\alpha,\beta)=\diy \int_0^\infty t^{\alpha-1}(1+t)^{-\alpha-\beta}\rd t,\quad \alpha>0,\;\beta>0.\eeqq
Similarly one yields
\beq\label{eq2:2.10} \eta_{\vp,p}(\mu)=\diy\frac{n}{2\mu+n+2}\Big\{\Psi\big(\mu+1\big)-\Psi\big(\mu+n/2+2\big)\Big\},\quad \mu>-1.\eeq
Further, consider $\vp(\bx)=\log \bx^T\bx$, then for $\mu>-1$, straightforwardly we can write
\beqq \eta_{\vp,p}(\mu)=\Big(\Psi\big(\mu+1\big)-\Psi\big(\mu+n/2+1\big)\Big)\Big(\Psi\big(n/2\big)-
\Psi\big(\mu+n/2+1\big)\Big)-\Psi'\big(\mu+n/2+1\big).\eeqq
Here $\Psi'$ stands the derivative function of $\Psi$.

\begin{thm}
Suppose $f$ is a PDF and $\bx\in\bbR^n\mapsto\vp(\bx)\geq 0$. Then $g_{p,\BC}$ is the unique maximizer of the WE $h^{\rm w}_{\vp}(f)$ under constrains
\beq \begin{array}{l} \quad\diy\bbE_{g_{p,\BC}}[\vp]\leq \bbE_f[\vp], \quad \hbox{and}\\
\\
\diy\int_{\bbS_{p,\BC}}\vp(\bx)\;f(\bx)\log\Big(1+(1-p)\beta \bx^T\BC^{-1}\bx\Big)\;\rd \bx\\
\quad\quad \diy \leq \eta^*_{\rho^*,p}(\frac{1}{1-p})+(1-p)\log A_p\Big(\bbE_f[\vp]-\bbE_{g_{p,\BC}}[\vp]\Big),\;\; \hbox{if}\;\; p<1,\\
\quad \quad\diy \geq \eta_{\rho,p}(\frac{1}{p-1})+(1-p)\log A_p\Big(\bbE_f[\vp]-\bbE_{g_{p,\BC}}[\vp]\Big),\;\;\hbox{if}\;\;p>1.\end{array}\eeq
with equality $f\equiv g_{p,\BC}$. Here $\rho^*$, $\rho$ form as (\ref{Def:1.12}) and $\eta^*$, $\eta$ stands as before, (\ref{def:eta}).
\end{thm}

The next step is to recall Theorem 6. from \cite{CHV}, omitting the proof.
\def\BI{\mathbf{I}} \def\BZ{\mathbf{Z}}\def\bz{\mathbf{z}}
\begin{thm}
Let $\BX$, $\BY$ be two independent RVs with covariance matrices $\BC_{\BX}=\BC_{\BY}=\BI_n$ and odd degrees of freedom $p_{\bx}$, $p_{\by}$ and PDFs $g_{p_{\bx},\BI}$, $g_{p_{\by},\BI}$ respectively, where
$$g_{p_{\bx},\BC}(\bx)=\big(p_{\bx}-2\big)^{-n/2}\;g_{p,\BC}\big((p_{\bx}-2)^{-1/2}\bx\big),\quad p=\diy\frac{p_{\bx}+n-2}{p_{\bx}+n}.$$
Then for $0\leq \lambda\leq 1$, the distribution of $\BZ=\lambda\;\BX+(1-\lambda)\BY$ is
\beq g_{\BZ}(\bz)=\sum\limits_{k=0}^{p_{\bz}} \alpha_k \;g_{p_{2k+1},\BI}(\bz).\eeq
where $p_{\bz}\leq \diy\frac{p_{\bx}+p_{\by}}{2}-1$. Note that throughout the note we shall use $g_p$ instead of $g_{p,\BI}$ as well.
\end{thm}
\begin{thm}
Suppose that $\BX$, $\BY$ are two independent RVs with PDFs $g_{p_{\bx}}$, $g_{p_{\by}}$, such that $\BC_{\BX}=\BC_{\BY}=\BI_n$ and $p_{\bx}$, $p_{\by}$ are odd freedom degrees. Assume RV $\BZ=\frac{\BX+\BY}{2}$. Set
\beq\label{Def:Delta}
\Delta_n(p)=\diy\Psi\Big(\frac{p+n}{2}\Big)-\Psi\Big(\frac{p}{2}\Big).\eeq
Then regarding the weighted Kullback-Leibler divergence the distribution $g_{p^*}$ with freedom degree $p^*$ which obeys
\beq\label{closePDF}
\Delta_n(p^*)=\diy \bbE_{\BW}\bigg(\bbE_{g_{p_{2k+1}}}
\Big[\vp\;\log(1+\BZ^T\BZ)\Big]\bigg)\Big/\bbE_{\BW}\Big(\bbE_{g_{p_{2k+1}}}[\vp]\Big),\eeq
or equivalently
\beq\label{Equi:closePDF} \diy \bbE_{g_{p^*}}\Big[\log\big(1+\BX^T\BX\big)\Big]\;\bbE_{g_{\BZ}}\Big[\vp\Big]=\bbE_{g_{\BZ}}\Big[\vp(\BX)\log \big(1+\BX^T\BX\big)\Big],\eeq
is the closest to the distribution $\BZ$, $g_{\BZ}$. Here the RV $\BW$ is distributed as
\beq \label{PDF:W} P\Big(\BW=2k+1\Big)=\alpha_k.\eeq
\end{thm}
\def\BW{\mathbf{W}}
{\bf Proof:} Taking into account (\ref{eq:1.2}), observe that for $g_{p^*}$ one can write
\beq\label{eq2:2.12} D^{\rm w}_{\phi}(g_{\BZ}\|g_{p^*})=-h^{\rm w}_{\vp}(g_{\BZ})-\int_{\bbR^n} \vp(\bz)\;g_{\BZ}(\bz)\log\;g_{p^*}(\bz)\rd \bz. \eeq
In order to explore the optimal value for $p^*$ we minimizes (\ref{eq2:2.12}). This is equivalent to find the maximizer of the above integral. For this reason we focus on the $\diy\int_{\bbR^n}\vp(\bz) g_{\BZ}(\bz)\log\;g_{p^*}(\bz)\rd \bz$. Thus one derives
\beq\label{eq2:2.13} \begin{array}{ccl}
\diy\int_{\bbR^n} \vp(\bz)\;g_{\BZ}(\bz)\log\;g_{p^*}(\bz)\rd \bz&=&\diy \sum\limits_{k=0}^{p_{\bz}}\alpha_k\int_{\bbR^n}\vp(\bz)\; g_{p_{2k+1}}(\bz)\log\;g_{p^*}(\bz)\;\rd \bz\\
&=&\diy\Big(\log A'_{p^*}\Big)\;\sum\limits_{k=0}^{p_{\BZ}}\alpha_k\int_{\bbR^n}\vp(\bz)\; g_{p_{2k+1}}(\bz)\;\rd \bz\\
&+&\diy\sum\limits_{k=0}^{p_{\BZ}}\alpha_k\int_{\bbR^n}\vp(\bz)\; g_{p_{2k+1}}(\bz)\log\;\Big(1+\bz^T\bz\Big)^{-(p^*+n)/2}\rd\bz.\end{array}\eeq
Substituting
$$A'_{p^*}=\diy\Gamma\Big(\frac{p^*+n}{2}\Big)\big/ \Gamma\Big(\frac{1}{2}\Big)\Gamma\Big(\frac{p^*}{2}\Big)$$
 in expression (\ref{eq2:2.13}), the LHS turns into
\beq \label{Eq2:2.14}\sum\limits_{k=0}^{p_{\bz}}\alpha_k\;\bbE_{g_{p_{2k+1}}}[\vp]\log \frac{\Gamma\Big(\frac{p^*+n}{2}\Big)}{\Gamma\Big(\frac{1}{2}\Big)\Gamma\Big(\frac{p^*}{2}\Big)}-
\frac{p^*+n}{2}\sum\limits_{k=0}^{p_{\bz}}\alpha_k\bbE_{g_{p_{2k+1}}}\Big[\vp\;\log(1+\BZ^T\BZ)\Big].\eeq
Taking the derivative of (\ref{Eq2:2.14}) with respect to $p^*$ obtains
\beq \label{eq2:2.16} \diy\frac{\Delta_n(p^*)}{2}\;\bbE_{\BW}\Big(\bbE_{g_{p_{2k+1}}}[\vp]\Big)-\frac{1}{2}\bbE_{\BW}\bigg(\bbE_{g_{p_{2k+1}}}
\Big[\vp\;\log(1+\BZ^T\BZ)\Big]\bigg),\eeq
where $\Delta_n(p^*)$ reads (\ref{Def:Delta}) and $\BW$ is denoted for a RV with distribution according to (\ref{PDF:W}). Equating (\ref{eq2:2.16}) to zero the desired result is achieved. Now it only remains to check the second derivative of (\ref{eq2:2.16}). We know that the derivative function of $\Psi$ is non-increasing, thus
\beqq \frac{\partial}{\partial p}\Delta_n(p)=\frac{1}{2} \Psi'\Big(\frac{p+n}{2}\Big)-\frac{1}{2} \Psi'\Big(\frac{p}{2}\Big)\leq 0.\eeqq
In addition, following the arguments in \cite{Z}, \cite{CHV}, one has
\beqq \Delta_n(p^*)=\bbE_{g_{p^*}}\Big[\log\;\big(1+\BX^T\BX\big)\Big].\eeqq
This together with
\beqq \begin{array}{ccl}\diy\bbE_{\BW}\bigg(\bbE_{g_{p_{2k+1}}}
\Big[\vp\;\log(1+\BZ^T\BZ)\Big]\bigg)&=& \diy\int_{\bbR^n} \sum\limits_{k=0}^{p_{\BZ}}\alpha_k\;g_{p_{2k+1}}(\bz)\; \vp(\bz)\log\;\big(1+\bz^T\bz)\;\rd \bz\\
\diy&=&\bbE_{g_{\BZ}}\Big[\vp(\BZ)\log\big(1+\BZ^T\BZ\big)\Big],\end{array}\eeqq
and
\beqq \bbE_{\BW}\Big(\bbE_{g_{p_{2k+1}}}[\vp]\Big)=\diy\int_{\bbR^n} \sum\limits_{k=0}^{p_{\BZ}}\alpha_k\;g_{p_{2k+1}}(\bz)\; \vp(\bz)\;\rd \bz=\bbE_{g_{\BZ}}\big[\vp(\BZ)\big].\eeqq
leads directly to (\ref{Equi:closePDF}). $\quad$ $\blacksquare$

\vspace{0.5 cm}
In the remaining arguments of this section, we shall address the reader to the following lemma as a technical low bound for the WRE.

\begin{lem}
Assume the sequence of PDFs $f_1,\dots,f_m$ on $\bbR^n$. Consider constants $s_1,\dots,s_m$ such that $\diy\sum\limits_{i=1}^m s_i=1$. For the mixture PDF $f$,
\beqq f(\bx)=\diy\sum\limits_{i=1}^m s_i\;f_i(\bx), \;\;\bx\in\bbR^n,\eeqq
and given WF $\vp$, we have
\beq h^{\rm w}_{\vp,p}(f)\geq \min\limits_{1\leq i\leq m} h^{\rm w}_{\vp,p}(f_i).\eeq
\end{lem}

{\bf Proof}: We begin with case $0<p<1$ which holds because of the concavity property for the WRE. Next suppose that $p>1$, then we can write
\beqq \begin{array}{ccl}
\diy\log\int_{\bbR^n} \vp(\bx)\;f^p(\bx)\;\rd \bx&=&\diy \log\int_{\bbR^n}\vp(\bx)\bigg(\sum\limits_i s_i\;f_i(\bx)\bigg)^p\rd\bx\\
&\leq& \diy\log \sum\limits_i s_i\;\int_{\bbR^n}\vp(\bx)  f_i^p(\bx)\;\rd \bx\\
&\leq& \diy \log \max\limits_{1\leq i\leq m}\diy\int_{\bbR^n}\vp(\bx) f_i^p(\bx)\;\rd \bx\\
&=&\diy\max\limits_{1\leq i\leq m}\log\diy\int_{\bbR^n}\vp(\bx)\;f_i^p(\bx)\;\rd \bx. \end{array}\eeqq
The proof of theorem is completed by observing that $\frac{1}{1-p}$ is negative. $\quad$ $\blacksquare$

\section{Extended Hadamard inequality and its consequences}
In this section, our aim is to establish a generalized form for Hadamard inequality in terms of WRE for given WF $\vp$. Combining this with maximum WRE distributions leads us to the assertions claimed in Corollary \ref{cor1}, \ref{cor2}. The following lemma is an immediate application of Lemma \ref{lem1}.
\begin{lem}\label{lem2}
Let $\BX$ be a RV with PDF $f$ with components $X_i;\Omega\mapsto\bbR$, $1\leq i\leq n$ having marginal PDF $f_i$, and joint PDF $f$. Given the WFs $\vp_i\geq 0$, such that $\vp(\bx)=\prod\limits_{i=1}^n\vp_i(x_i)$ is considered as the WF. Suppose that
\beq \diy\int_{\bbR^n}\prod\limits_{i=1}^n\vp_i(x_i)f^{p-1}_i(x_i)\;\bigg[f(\bx)-\prod\limits_{i=1}^n f_i(x_i)\bigg]\rd \bx\leq (\geq) 0,\quad \hbox{for}\;\;0<p<1\;(p>1).\eeq
Then
\beq\label{Inq:lem2} h^{\rm w}_{\vp,p}(f)\leq \sum\limits_{i=1}^n h^{\rm w}_{\vp_i,p}(f_i).\eeq
The equality here holds iff the components $X_1,\dots,X_n$ are independent.
\end{lem}

\begin{thm}\label{extend:Hadamard}
{\rm (The extended Hadamard inequality, Theorem 3.9 cf. \cite{SY}).} Let $\BC=(C_{ij})$ be a positive definite $n\times n$ matrix and $g_{p,\BC}$ stands the PDF in Definition \ref{def:gc}. In addition let $g_{p,C_{ii}}$ be the marginal PDF of the $i$-th components, that is as in (\ref{def.gc}) when $n=1$. Then for given functions $x_i\in\bbR \mapsto\vp_i(x_i)\geq 0$, $1\geq i\geq n$ which if $n/(n+2)<p<1\;(p>1)$ obey
\beq \diy\int_{\bbS_{p,\BC}} \prod\limits_{i=1}^n \vp_i(x_i)g^{p-1}_{p,C_{ii}}(x_i)\;\bigg[g_{p,\BC}(\bx)- \prod\limits_{i=1}^n g_{p,C_{ii}}\bigg]\rd \bx\leq(\geq) 0,\eeq
one has, $n/(n+2)<p<1$,
\beq\label{Inq:Hadamard}\begin{array}{l} \diy \frac{1-p}{2}\log \prod\limits_iC_{ii}+\sum\limits_i \log \varpi^*_1(p)\;\alpha^*_{\vp_i,p}(C_{ii})\\
\qquad \quad -\diy \frac{1-p}{2}\log {\rm det}\;\BC-\log \varpi^*_n(p)\;\alpha^*_{\vp,p}(\BC)\geq 0.\end{array}\eeq
Here $\varpi^*_n(p)$, $\alpha^*_{\vp,p}(\BC)$ are as in (\ref{varpi*}),(\ref{Def:1.13}) and
\beq\label{alpha*ii}\alpha^*_{\vp_i,p}(C_{ii})=\diy \bbE\;\left[\vp_i\bigg(\Big\{C_{ii}\big((3p-1)/(1-p)\big)\Big\}^{1/2} Y^*_i\bigg)\right],\quad p\in(1/3,1).\eeq
where $Y^*_i$ has the Pearson's type VII univariate distribution with parameter $\mu=p/(1-p)$.\\
For case $p>1$, recall (\ref{varpi}), (\ref{Def:1.13}) and in (\ref{Inq:Hadamard}) swap $\varpi_n(p)$, $\alpha_{\vp,p}(\BC)$, $\alpha_{\vp_i,p}(C_{ii})$ with $\varpi^*_n(p)$, $\alpha^*_{\vp,p}(\BC)$, $\alpha^*_{\vp_i,p}(C_{ii})$. Note that $\alpha_{\vp_i,p}(C_{ii})$ is defined in similar manner as (\ref{alpha*ii}) by replacing random variable $Y_i$ in $Y^*_i$ where $Y_i$ has the Pearson's type II univariate distribution with parameter $p/(p-1)$. The equality in (\ref{Inq:Hadamard}) occurs iff $\BC$ is diagonal.
\end{thm}

{\bf Proof:} We give the proof for part $n/(n+2)<p<1$, while the proof for the case $p>1$ follows in a similar manner. Assume that RV $\BX$ has PDF $g_{p,\BC}$. By virtue of (\ref{Inq:lem2}), (\ref{WREgc}) one yields
\beqq \begin{array}{l}
\diy\frac{1}{1-p}\log \varpi^*_{n}(p)+\frac{1}{2}\log {\rm det}\;\BC+\frac{1}{1-p}\log \alpha^*_{\vp,p}(\BC)\\
\qquad\quad \leq \sum\limits_{i=1}^n \bigg(\frac{1}{1-p}\log \varpi^*_1(p)+\frac{1}{2}\log C_{ii}+\frac{1}{1-p}\log \alpha^*_{\vp_i,p}(C_{ii})\bigg).\end{array}\eeqq
Here quantity $\alpha^*_{\vp_i,p}(C_{ii})$ is introduced by (\ref{alpha*ii}). The assertion (\ref{Inq:Hadamard}) then follows. Note that the case equality is covered by the equality in Lemma \ref{lem2}. $\quad$ $\blacksquare$
\vskip 0.5 truecm
Now, we recall the following definition from \cite{SPK}, which is essentially the integral representation of the modified Bessel function of the third and first kinds. ( see \cite{Wa}, p. 182).
\begin{defn}\label{defn:2}
{\rm (a)} The integral representation of the modified Bessel function of the third kind, denoted as $K_{\lambda}(z)$, is defined by
\beq K_{\lambda}(z)=\diy\frac{1}{2} \int_0^\infty x^{\lambda-1}\exp\Big\{-\frac{1}{2} z\big(x+\frac{1}{x}\big)\Big\}\rd x, \quad z>0.\eeq
Note that $K_{\lambda}(z)=K_{-\lambda}(z), \;z>0,\; \lambda\in\bbR$ and
$$K_{\lambda}(z)\cong \Gamma(\lambda)2^{\lambda-1}z^{-\lambda},\;\; \hbox{as}\; z\rightarrow 0^+,\;\lambda>0.$$
{\rm (b)} The Bessel function of the first kind, written as $J_{\gamma}(z)$, is given in form
\beq \label{Bessel.first.kind} J_{\gamma}(z)=\diy\frac{1}{\pi}\int_0^\pi\cos\big(z\sin\theta-n\theta\big)\rd \theta.\eeq
\end{defn}
The techniques developed by the WRE in Theorem \ref{extend:Hadamard} so far allow us to establish Corollary \ref{cor1} below rendering a general form of Hadamard inequality. To this end, we firstly introduce more notations:
\beq \label{def:2.14}\epsilon_n=2\Big(\frac{p}{1-p}-\frac{n}{2}\Big)\quad \hbox{and}\quad \chi^*_n(p)=\diy\frac{\varpi^*_n(p)}{\Gamma(\epsilon_n/2)\;2^{\epsilon_n/2-1}},\;\; n/(n+2)<p<1.\eeq
In (\ref{def:2.14}), by setting $n=1$ we get $ \epsilon_1$ and $\chi^*_1(p)$. Also set
\beq \label{def2:2.14} \xi_n=2\Big(\frac{p}{p-1}+\frac{n}{2}\Big)\quad \hbox{and}\quad \chi_n(p)=2^{\xi_n/2}\Gamma(\xi_n/2+1) \varpi_n(p),\;\; p>1.\eeq
Observe that $\xi_1$ and $\chi_1(p)$ are obtained if in (\ref{def2:2.14}) we consider $n=1$.
\def\bt{\mathbf{t}}
\begin{cor}\label{cor1}
Suppose $\BC$ is a positive definite $n\times n$ matrix. Consider the vector $\bt=(t_1,\dots,t_n)\in\bbR^n$ is satisfied in
\beq \diy\int_{\bbS_{p,\BC}} e^{i\bt\bx}\prod\limits_{i=1}^n g^{p-1}_{p,C_{ii}}(x_i)\;\bigg[g_{p,\BC}(\bx)- \prod\limits_{i=1}^n g_{p,C_{ii}}\bigg]\rd \bx\leq (\geq) 0,\;\; n/(n+2)<p<1(p>1)\eeq
Invoking $K_{\lambda}(z)$ in Definition \ref{defn:2} gives the bound
\beq \begin{array}{l}
\diy\frac{1-p}{2}\log \prod\limits_{i}C_{ii}+\sum\limits_{i}\log \bigg(\chi^*_1(p)\;K_{\epsilon_1/2}\Big(\sqrt{\epsilon_1C_{ii}}\;|t_i|\Big)\Big(\sqrt{\epsilon_1C_{ii}}\;|t_i|\Big)^{\epsilon_1/2}\bigg)\\
\diy -\frac{1-p}{2}\log {\rm det}\;\BC-\log \bigg(\chi^*_n(p)\;K_{\epsilon_n/2}\Big(\|\sqrt{\epsilon_n}\;\BC^{1/2}\bt\|\Big)\Big(\|\sqrt{\epsilon_n}\;\BC^{1/2}\bt\|\Big)^{\epsilon_n/2}\bigg)\geq 0,\end{array}\eeq
when $n/(n+2)<p<1$. One can also for $p>1$ deduce
\beq \begin{array}{l}
\diy\frac{1-p}{2}\log \prod\limits_{i}C_{ii}+\sum\limits_{i}\log \bigg(\chi_1(p)\;J_{\xi_1/2}\Big(\sqrt{\xi_1C_{ii}}\;|t_i|\Big)\Big(\sqrt{\xi_1C_{ii}}\;|t_i|\Big)^{\xi_1/2}\bigg)\\
\diy -\frac{1-p}{2}\log {\rm det}\;\BC-\log \bigg(\chi_n(p)\;J_{\xi_n/2}\Big(\|\sqrt{\xi_n}\;\BC^{1/2}\bt\|\Big)\Big(\|\sqrt{\xi_n}\;\BC^{1/2}\bt\|\Big)^{\xi_n/2}\bigg)\geq 0,\end{array}\eeq
Here $J_.$ refers to the Bessel function of the first kind indicated in (\ref{Bessel.first.kind}). Further particularly, for a positive definite $2\times 2$ matrix one has
\beq\label{Ineq:2.17} \diy\frac{1}{6}\log \frac{C_{11}C_{22}}{{\rm det}\;\BC}+\log \frac{\prod\limits_{i=1}^2 K_{3/2}\Big(\sqrt{3C_{ii}}\;|t_i|\Big)\Big(\sqrt{3C_{ii}}\;|t_i|\Big)^{3/2}}{K_1\Big(\|\sqrt{2}\;\BC^{1/2}\bt\|\Big)\Big(\|\sqrt{2}\;\BC^{1/2}\bt\|\Big)}\geq
\log \frac{3\;\pi^{2/3}}{4}.\eeq
\end{cor}

{\bf Proof:} The proof for part $n/(n+2)<p<1$ is provided. By using $\diy\vp_l(x_l)=e^{it_lx_l}$ in (\ref{Def:1.13}), for $\BY^*\sim f_{VII}$ and $\bt\in\bbR^n$, we have
\beqq\alpha^*_{\vp,p}(\BC)=\bbE_{f_{VII}}\bigg[\prod\limits_{l=1}^n e^{i\sqrt{\epsilon_n} t_l\sum\limits_j Y^*_j C_{lj}^{1/2}}\bigg]=\phi_{\BY^*}(\sqrt{\epsilon_n}\BC^{1/2}\bt),\eeqq
where $\epsilon_n$ stands as in (\ref{def:2.14}) and $\phi_{\BY^*}$ represents the characteristic function for RV $\BY^*$. By virtue of Result 4, cf. \cite{SPK} one yields
\beq \label{eq:2.18}\alpha^*_{\vp,p}(\BC)=\diy\frac{K_{\epsilon_n/2}\Big(\|\sqrt{\epsilon_n}\;\BC^{1/2}\bt\|\Big)
\Big(\|\sqrt{\epsilon_n}\;\BC^{1/2}\bt\|\Big)^{\epsilon_n/2}}{\Gamma(\epsilon_n/2)2^{\epsilon_n/2-1}}.\eeq
Similarly, in accordance with (\ref{alpha*ii}) one can derive
\beq\label{eq:2.19}\begin{array}{l}\diy \alpha^*_{\vp_l,p}(C_{ll})=\bbE\Big[e^{i \sqrt{\epsilon_1C_{ll}}\;t_lY^*_l}\Big]
=\diy\phi_{Y^*_l}\big(\sqrt{\epsilon_1C_{ll}}\;t_l\big)\\
\quad\quad=\diy\frac{K_{\epsilon_1/2}\Big(\sqrt{\epsilon_1C_{ll}}\;|t_l|\Big)
\Big(\sqrt{\epsilon_1C_{ll}}\;|t_l|\Big)^{\epsilon_1/2}}{\Gamma(\epsilon_1/2)2^{\epsilon_1/2-1}}.\end{array}\eeq
Replace (\ref{eq:2.18}), (\ref{eq:2.19}) in (\ref{Inq:Hadamard}). This concludes the proof. The assertion (\ref{Ineq:2.17}) is achieved by choosing $p=\frac{2}{3}$, $n=2$. $\quad$ $\blacksquare$\\

In Lemma \ref{lem:3} we extend the results of Lemma \ref{lem2} to exponents of WREs for sub-strings $(\BX_1,\BX_2)$ in $\BX$ having $g_{p,\BC}$ PDF. We verify this by owing to Theorem 3 in \cite{CHV} and Lemma \ref{lem1} straightforwardly, hence the proof is omitted.
\begin{lem}\label{lem:3}
Let $\BX^T=(\BX_1^T,\BX_2^T)$ be a RV in $\bbR^n$, with PDF $g_{p,\BC}$ and characteristic matrix $\BC=\big(\BC_{ij}\big)$, $i,j=1,2$, where ${\rm dim}\;\BX_i=n_i$, $n_1+n_2=n$ and ${\rm dim}\;\BC_{ij}=n_i\times n_j$. Then for given WF $\vp(\bx)=\prod\limits_{j=1,2}\vp_j(\bx_j)$ which for $0<p<1\; (p>1)$ is satisfied in below
\beqq \diy\int_{\bbR^n} \prod\limits_{j=1,2}\vp_j(\bx_j) g_{p_j,\BC_{jj}}^{p-1}(\bx_j)\;\Big[g_{p,\BC}(\bx)-\prod\limits_{j=1,2}g_{p_j,\BC_{jj}}(\bx_j)\Big]\;\rd \bx\leq(\geq)0.\eeqq
where index $p_j$ is given by
$$\diy\frac{1}{1-p_j}=\frac{1}{1-p}-\frac{n_j}{2},\;\; \hbox{and}\;\; \BX_j\sim g_{p_j,\BC_{jj}}.$$
we have
\beq h^{\rm w}_{\vp,p}(\BX)\leq h^{\rm w}_{\vp_1,p}(\BX_1)+h^{\rm w}_{\vp_2,p}(\BX_2).\eeq
\end{lem}
\def\BB{\mathbf{B}}
Next step would be to use Lemma \ref{lem:3} and explore an upper bound for determinant of block matrices in terms of the expected value for the WF $\vp$, i.e. $\alpha_{\vp,p}$.
\begin{thm}\label{thm:matrix}
Consider block matrix $\BB=\big(\BB_{ij}\big)$, $i,j=1,2$ with ${\rm dim}\;\BB_{ij}=n'_i\times n_j$, $n=n_1+n_2$, $n'=n'_1+n'_2$. Furthermore let $\BC=\big(\BC_{ij}\big)$, be positive definite block matrix where ${\rm dim}\;\BC_{ij}=n_i\times n_j$ therefore ${\rm dim}\;\BC=n\times n$. Assume that $p',\; p'_1,\;p'_2$ follow relation
\beq\label{def:p's} \diy\frac{1}{1-p'_i}=\frac{1}{1-p'}-\frac{n'_i}{2},\quad\hbox{whereas}\;\; p'>\diy\frac{(1-p'_i)\;n'_i}{2},\;\;i=1,2\eeq
and take ranges
\beqq p'\in\big(\frac{n'}{n'+2},1\big)\quad \hbox{and}\quad p'_i\in\big(\frac{n'_i}{n'_i+2},1\big),\;\; i=1,2.\eeqq
Define
\beqq \zeta(p',p'_1,p'_2)=\diy\frac{\varpi^*_{n'_1}(p'_1,p')\;\varpi^*_{n'_2}(p'_2,p')}{\varpi^*_{n'}(p')},\eeqq
where $\varpi^*$s denote the corresponding quantities in (\ref{varpi*}). Now by recalling $\alpha^*$ from (\ref{Def:1.12}) and the given function $\vp=\prod\limits_{i=1,2}\vp_i$, $\vp_i\geq 0$, if
\beq \diy\int_{\bbR^{n'}} \prod\limits_{j=1,2}\vp_j(\bx_j) g_{p'_j,\BC_{jj}}^{p'-1}(\bx_j)\;\Big[g_{p',\BB\BC\BB}(\bx)-\prod\limits_{j=1,2}g_{p'_j,\BC'_{j}}(\bx_j)\Big]\;\rd \bx\leq(\geq)0.\eeq
holds true then the following inequality is emerged:
\beq\label{main:ineq} \bigg(\diy\frac{{\rm det}\; \BB\BC\BB^T}{\big({\rm det}\;\BC'_1\big)\;\big({\rm det}\;\BC'_2\big)}\bigg)^{1-p'}\bigg(\diy\frac{\alpha^*_{\vp,p'}(\BB\BC\BB^T)}{\alpha^*_{\vp_1,p'_1}(\BC'_1)\;\alpha^*_{\vp_2,p'_2}(\BC'_2)}\bigg)^2\leq \zeta(p',p'_1,p'_2).\eeq
Here $\BC'_1$, $\BC'_2$ with ${\rm dim}\;\BC'_i=n'_i\times n'_i$, $i=1,2$ represent the diagonal block matrices $\BB\BC\BB^T$, that is
\beqq\begin{array}{cl}
\BC'_1=\diy\sum\limits_{i=1,2} \BB_{1i}\BC_{ii}\BB^T_{1i}+\BB_{12}\BC_{21}\BB^T_{11}+\BB_{11}\BC_{12}\BB^T_{12},\\
\BC'_2=\diy\sum\limits_{i=1,2} \BB_{2i}\BC_{ii}\BB^T_{2i}+\BB_{22}\BC_{21}\BB^T_{21}+\BB_{21}\BC_{12}\BB^T_{22}.\end{array}\eeqq
\end{thm}

{\bf Proof:} Let $\BX_1$ and $\BX_2$ be $n_1$ and $n_2$ RVs mutually distributed according to the Renyi entropy maximizing density $g_{p,\BC}$, where
$$\diy\frac{1}{1-p}=\diy\frac{1}{1-p'}+\frac{n-n'}{2}.$$
Consider RV $\BX^T=(\BX_1^T,\BX_2^T)$ and deduced RV $\BY=\BB\;\BX=\BB\left[\begin{array}{c} \BX_1\\ \BX_2\end{array}\right]=\left[\begin{array}{c} \BY_1\\ \BY_2 \end{array}\right]$, with $\BY_i=\BB_{i1}\BX_1+\BB_{i2}\BX_2$, $i=1,2$. Owing to the Theorem 4, \cite{CHV}, the RV $\BY$ follows $g_{p',\BC'}$ distribution with characteristic matrix $\BC'=\BB\BC\BB^T$, with $p'$ given by (\ref{def:p's}). Next by virtue of the Theorem 3 from \cite{CHV}, we can derive that the marginal density of RV $\BY_i$ is $g_{p'_i,\BC'_ii}$ such that ${\rm dim}\;\BC'_{ii}=n'_i\times n'_j$ and $p'_i$ comes again from (\ref{def:p's}). In this stage recall the Lemma \ref{lem:3}, therefore one yields
\beqq h^{\rm w}_{\vp,p'}(\BY)\leq h^{\rm w}_{\vp_1,p'}(\BY_1)+  h^{\rm w}_{\vp_2,p'}(\BY_2).\eeqq
Equivalently
\beqq \begin{array}{l}
\diy\frac{1}{1-p'}\log\;\varpi^*_{n'}(p')+\frac{1}{2}\log \; {\det}\;\BC'+\frac{1}{1-p'}\log\;\alpha^*_{\vp,p'}(\BC')\\
\quad\leq\diy \frac{1}{1-p'}\log\;\varpi^*_{n'_1}(p'_1,p')+\frac{1}{2}\log \; {\det}\;\BC'_{11}+\frac{1}{1-p'}\log\;\alpha^*_{\vp_1,p'_1}(\BC'_{11})\\
\quad+\diy\frac{1}{1-p'}\log\;\varpi^*_{n'_2}(p'_2,p')+\frac{1}{2}\log \; {\det}\;\BC'_{22}+\frac{1}{1-p'}\log\;\alpha^*_{\vp_2,p'_2}(\BC'_{22}).\end{array}\eeqq
Finally, after direct computations and inserting the constant $\zeta(p',p'_1,p'_2)$, the property claimed in (\ref{main:ineq}) is obtained. $\quad$ $\blacksquare$

\vspace{0.5 cm}
\def\BL{\underline{\lambda}}\def\BI{\mathbf{I}}
\def\BBL{\mathbf{\Lambda}}
\def\bI{\mathbf{1}}\def\B0{\mathbf{0}}
As closing, the inequality (\ref{main:ineq}) is analyzed for particular case of $\BC$ and $\BB$. So we offer
\begin{cor}\label{cor2}
Let $\BY^*_1$ and $\BY^*_2$ be $n'_1$ and $n'_2$ RVs, following Pearson's type VII with parameters $\frac{p'}{1-p'_1}$ and $\frac{p'}{1-p'_2}$ respectively. Further suppose that $n'=n'_1+n'_2$ RV ${\BY^*}^T=({\BY^*_1}^T,{\BY^*_2}^T)$ has Pearson's type VII with parameter $\frac{p'}{1-p'}$ as well. Here $p',\;p'_1,\;p'_2$ are as in Theorem \ref{thm:matrix}. Next let $\BL^T=(\BL_1^T,\BL_2^T)$ be $n'$ RV such that $\BL_1^T=(\lambda_1,\dots \lambda_{n'_1})$, $\BL_2^T=(\lambda_{n'_{1}+1},\dots,\lambda_{n'})$ and $\lambda_i\geq 0$. Consider $\BBL_1$ and $\BBL_2$ are the diagonal matrices with diagonal vectors $\BL_1$ and $\BL_2$ respectively, set
$$\BBL=\left(\begin{array}{cc} \BBL_1& \B0\\ \B0 & \BBL_2\\ \end{array}\right).$$
Assume the following inequality is satisfied:
\beq \diy\int_{\bbR^{n'}}\prod\limits_{i=1,2}|\bx_i|\;g^{p'-1}_{p'_{i},\bI}(\bx_i)\;\Big[g_{p',\BBL}(\bx)
-\prod\limits_{i=1,2}g_{p'_i,\BBL_i}(\bx_i)\Big]\;\rd\bx\leq0, \quad \bx_i\in\bbR^{n'_i},\;i=1,2.\eeq
Then
\beq\label{cor1} \diy\bbE^2\Big[|\BL^T\BY^*|\Big]\leq \diy\eta(\BL)\;\bbE^2\Big[|\BL_1^T\BY^*_1|\Big]\;\bbE^2\Big[|\BL_2^T\BY^*_2|\Big],\eeq
where
\beqq \eta(\BL)=\left(\diy\frac{\sum\limits_{i=1}^{n'}\lambda_i^2}{\Big(\sum\limits_{i=1}^{n'_1}\lambda_i^2\Big)\;
\Big(\sum\limits_{i=n'_1+1}^{n'}\lambda_i^2\Big)}\right)^{p'-1}\;\zeta(p',p'_1,p'_2)\;
\left(\frac{2\Big(\frac{p'_1}{1-p'_1}-\frac{n'_1}{2}\Big)\Big(\frac{p'_2}{1-p'_2}-\frac{n'_2}{2}\Big)}
{\Big(\frac{p'}{1-p'}-\frac{n'}{2}\Big)}\right).\eeqq
\end{cor}

{\bf Proof:} Suppose that $\vp_i(\bx)=|\bx|$, $i=1,2$ The assertion (\ref{cor1}) directly can be proved by choosing diagonal matrix $\BB$ with entries $\lambda_i$, $i=1,\dots, n'$ and $\BC=\BI$ is considered identity matrix in Theorem \ref{thm:matrix}.   $\quad$ $\blacksquare$\\
\def\BA{\mathbf{A}} \def\BB{\mathbf{B}}\def\BU{\mathbf{U}} \def\BV{\mathbf{V}}\def\BD{\mathbf{D}}

Invoking \cite{JV}, for fixed $p>1$, consider two $n$-dimensional RVs $\BX$, $\BY$ with corresponding distributions $g_{p,\BC_{\BX}}$, $g_{p,\BC_{\BY}}$. Then the convolution $\BX*_p\BY$ defined below is also a Renyi entropy maximizer, having $g_{p,\BC_\BX+\BC_\BY}$ PDF:
\beqq \BX*_p\BY=\diy\frac{U_{\BX}\BX+U_{\BY}\BY}{\sqrt{\Big(U_{\BX}\BX+U_{\BY}\BY\Big)^T(m\BC_{\BX\BY})^{-1}
\Big(U_{\BX}\BX+U_{\BY}\BY\Big)+V}},\eeqq
where $\BC_{\BX\BY}=\BC_{\BX}+\BC_{\BY}$ and $U_{\BX},\;U_{\BY}\sim f_{m},\; m=n+2p/(p-1)$ and $V\sim f_{m}$ but $m=2p/(p-1)$ are independent random variables:
\beqq f_{m}(x)=\diy\frac{2^{1-m/2}}{\Gamma(m/2)} x^{m-1}\exp\Big(-\frac{x^2}{2}\Big), \;\; x>0.\eeqq
It is worthwhile to note that the convolution $\BX\circ\BY$ defined by
\beqq \BX\circ\BY=\Theta^{-1}_{(m-2)(\BC_\BX+\BC_\BY)}\bigg(\Theta_{(m-1)\BC_{\BX}}(\BX)*_{\widetilde{p}}\Theta_{(m-2)\BC_{\BY}}(\BY)\bigg),\eeqq
with $m=2/(1-p)-n$ and $\widetilde{p}$ satisfies in $1/(\widetilde{p}-1)=m/2-1$ and
\beqq \Theta^{-1}_\BD(\BX)=\diy\frac{\BX}{\sqrt{1-\BX^T\BD^{-1}\BX}},\eeqq
follows distribution $g_{p,\BC_\BX+\BC_\BY}$. The above properties promoted us focus on $g_{p,\BC_\BX}$ and $g_{\BC_\BX+\BC_\BY}$ which is encapsulated in the following theorem. Before, consider a WF $(\bx,\by)\in \bbR^n\times \bbR^n\mapsto \vp(\bx,\by)\geq 0$ and set
\beqq \vp^*_{\BC_{\BX}}(\bx)=\vp(\bx)\;\bx^T\BC_{\BX}^{-1}\bx.\eeqq
Note that here we have
\beq\label{eq:33:22}\bbE_{g_{p,\BC_{\BX}}}\big[\vp^*_{\BC_{\BX}}\big]={\rm tr}\;\BC_\BX^{-1}\Psi^g_{\BC_\BX}\;\;\; \hbox{and matrix}\;\Psi^g_{\BC_{\BX}}=\Big(\psi^{\BC_{\BX}}_{ij}\Big).\eeq
where
$$\psi^{\BC_{\BX}}_{ij}=\int_{\bbR^n}\vp(\bx)x_ix_j g_{p,\BC_{\BX}}(\bx)\;\rd \bx.$$
Moreover, similarly
\beq\label{eq:33:23}\bbE_{g_{p,\BC_{\BX}}}\big[\vp^*_{\BC_{\BX}+\BC_\BY}\big]={\rm tr}\; \big(\BC_\BX+\BC_\BY\big)^{-1} \Psi^g_{\BC_\BX}.\eeq
Now, owing to (\ref{eq:33:22}), (\ref{eq:33:23}) we provide a general result regarding the matrices, as stated by the following theorem.
\begin{thm}
Let $\BA$, $\BB$ be two $n\times n$ matrices. For given WF $\vp$ assume the following inequality involving $g_{p,\BA}$:
\beq\label{eq:3.22}\begin{array}{l}
\diy\Big({\rm det}\;(\BA+\BB)\Big)^{(1-p)/2}\bigg\{\bbE_{g_{p,\BA}}[\vp]+(1-p)\beta\bbE_{g_{p,\BA}}\big[\vp^*_{\BA+\BB}\big]\bigg\}\\
\quad \diy\leq (\geq) \diy\Big({\rm det}\;\BA\Big)^{(1-p)/2}\bigg\{\bbE_{g_{p,\BA}}[\vp]+(1-p)\beta\bbE_{g_{p,\BA}}\big[\vp^*_{\BA}\big]\bigg\},\;\; p<1\;(p>1).\end{array}\eeq
Then
\beq\label{eq:3.23}\bigg(\frac{\alpha^*_{\vp,p}(\BA+\BB)}{\alpha^*_{\alpha,p}(\BA)}\bigg)^{1/(1-p)}\geq \bigg(\frac{{\rm det}\;(\BA+\BB)}{{\rm det}\;\BA}\bigg)^{1/2},\;\; p<1.\eeq
For case $p>1$ substitute $\alpha_{\vp,p}$ in $\alpha^*_{\vp,p}$.
\end{thm}
\vskip .5 truecm

{\bf Proof:} We use the Renyi maximizing distributions with covariance matrices $\BA$ and $\BA+\BB$, i.e. $g_{p,\BA}$, $g_{p,\BA+\BB}$. Applying some straightforward computations, it can be seen that (\ref{eq:3.22}) implies
\beqq \diy\frac{1}{1-p}\log \bigg(\int_{\bbR^n}\vp g_{p,\BA+\BB}^{p-1}g_{p,\BA}\rd \bx\bigg)\leq \frac{1}{1-p}\log \bigg(\int_{\bbR^n}\vp g_{p,\BA}^{p}\;\rd\bx\bigg).\eeqq
By virtue of (\ref{eq:2.7}), we can write
\beqq h^{\rm w}_{\vp,p}(g_{p,\BA+\BB})\leq h^{\rm w}_{\vp,p}(g_{p,\BA}),\eeqq
which consequently by inserting (\ref{WREgc}) for $p<1$ and (\ref{WREgc2}) when $p>1$, we obtain (\ref{eq:3.23}).\\
\hfill$\quad$ $\blacksquare$\\
\vskip .5 truecm
{\bf Remark:} Recalling the Sherman-Morrison formula (see \cite{DB}, p. 161 and \cite {SuYSSt}), If $\BA$ and $\BA+\BB$ be nonsingular matrices where $\BB$ is a matrix of rank one. Let $\kappa={\rm tr}\;\big(\BB\BA^{-1}\big)$, $\BB_\BA=\BA^{-1}\BB\BA^{-1}$. Then $\kappa\neq -1$ and condition (\ref{eq:3.22}) turns into the following inequality:
\beqq\begin{array}{l} \diy \bigg(\Big({\rm det}\;(\BA+\BB)\Big)^{(1-p)/2}-\Big({\rm det}\;\BA\Big)^{(1-p)/2}\bigg)\Big((1-p)\beta+1\Big)\bbE_{g_{p,\BA}}[\vp]\\
\qquad \leq (\geq)\diy\frac{(1-p)\beta} {1+\kappa} \Big({\rm det}\;(\BA+\BB)\Big)^{(1-p)/2}\bbE_{g_{p,\BA}}\big[\vp^*_{\BB_\BA}\big],\;\;\; p<1\;(p>1).\end{array}\eeqq

\vskip .5 truecm
{\emph{Acknowledgements --}}
SYS thanks the CAPES PNPD-UFSCAR Foundation
for the financial support in the year 2014-5. SYS thanks
the Federal University of Sao Carlos, Department of Statistics, for hospitality during the year 2014-5.

\vskip .5 truecm
\bibliographystyle{plain}

 
\vspace{0.5cm}
\noindent Salimeh Yasaei Sekeh\\
is with the Statistics Department, DEs,\\
of Federal\;University\;of\;S$\tilde{\rm a}$o\;Carlos (UFSCar),\\
S$\tilde{\rm a}$o\;Paulo, Brazil\\
E-mail: sa$_{-}$yasaei@yahoo.com
\vskip 15pt

\end{document}